\documentclass[twocolumn,superscriptaddress,prl,showpacs]{revtex4}
\usepackage[letterpaper,left=0.7in,right=0.7in,top=0.90in,bottom=0.8in]{geometry}
\usepackage{amssymb}
\usepackage{amsmath}
\usepackage{graphicx}
\usepackage{subfigure}

\begin{document}

\title{
%\qsgw\  calculations of the interface resonance states at the Fe/MgO interface
%Spin transport mediated by states at the Fe/MgO interface: transport theory based on the \qsgw\  calculations.
%Spin transport mediated by states at the Fe/MgO interface:
%theory based on electronic structure calculated  within the \qsgw\   theory.
%Polarized electron transport across Fe/Mgo interfaces:
%theory based on electronic structure calculated  within the \qsgw\  approach.
%Electron correlations and  spin polarized transport mediated by states localized at the Fe/MgO [001] interface.
Brillouin zone spin filtering mechanism of enhanced TMR and correlation effects in Co(0001)/\emph{h}-BN/Co(0001) magnetic tunnel junction
}

\author{Sergey V. Faleev}
\email{svfaleev@us.ibm.com}
\affiliation{IBM Almaden Research Center, 650 Harry Road, San Jose, California 95120, USA}
\author{Stuart S. P. Parkin}

\affiliation{IBM Almaden Research Center, 650 Harry Road, San Jose, California 95120, USA}
\author{Oleg N. Mryasov}
%\affiliation{MINT Center, University of Alabama, P.O. Box 870209, Tuscaloosa, Alabama 35487, USA}
\affiliation{Physics and Astronomy, University of Alabama, Tuscaloosa, AL, 35487, USA}

\date{\today}

\begin{abstract}
The 'Brillouin zone spin filtering'  mechanism of enhanced tunneling magnetoresistance (TMR) is described for  magnetic tunnel junctions (MTJ) and studied on an example of the MTJ with hcp Co electrodes and hexagonal BN (h-BN) spacer.
Our calculations based on local density approximation of density functional theory (LDA-DFT) for  Co(0001)/h-BN/Co(0001)  MTJ predict high  TMR in this device  due to Brillouin zone filtering mechanism.
Owning to the specific complex band structure of the h-BN the spin-dependent tunneling conductance of the system is ultra-sensitive to small variations of the Fermi energy position inside the BN band gap. Doping of the BN and, consequentially, changing the Fermi energy position could lead to variation of the TMR  by several orders of magnitude. We show also that taking into account correlation effects on beyond DFT level is required to accurately describe position of the Fermi level and thus transport propertied of the system.
%As we show here critical property needed for high TMR in this MTJs is an alignment of the Co Fermi energy ($E_F$) very close to valence band maximum ($E_V$) of h-BN. DFT-LDA theory predicts $\Delta E_{FV} \equiv E_F - E_V \approx 0.2$ eV, while  $\Delta E_{FV}$ calculated by   more accurate Quasiparticle Self-consistent GW (QSGW) theory increases to 2.2 eV. Thus   p-doping  of the  h-BN is required to reduce $\Delta E_{FV}$ to practical range of below 0.1 eV to ensure high TMR (that theoretically in the ballistic transport limit can be as high as several orders of magnitude).
Our study suggests that new MTJ based on   hcp Co-Pt or Co-Pd disordered alloy electrodes and p-doped hexagonal BN spacer is a promising  candidate for the spin-transfer torque   magnetoresistive random-access memory (STT-MRAM).

%By varying the dopant concentration ($\Delta E_{FV}$) and h-BN   thickness one can independently regulate the TMR and resistance of the MTJ.
\end{abstract}
\maketitle
%\special{papersize=8.5 in, 11 in}

Theoretical prediction of high TMR in Fe/MgO/Fe MTJ due to so-called 'symmetry spin filtering' mechanism \cite{Butler01,Mathon01}   and its quick experimental verification \cite{Parkin04,Yuasa04} revolutionized the hard disk drive (HDD) industry during the last decade. On the other hand, continues progress in new areas of magnetic memory technology (for example, STT-MRAM) demands for a development of novel MTJ in which high value of  the volume type  magnetocrystalline anisotropy (MCA) of electrodes is as important as high TMR.  Standard nowadays Fe-based electrodes have low volume MCA due to high (cubic) symmetry and thus has to rely on interface anisotropy  which  cannot fully satisfy strict demands of new technology.

In present paper we suggest different mechanism of high TMR in magnetic tunnel junctions which we call the 'Brillouin zone filtering'  and   study this mechanism in details on a specific example of low symmetry (hexagonal) hcp-Co/h-BN/hcp-Co junction.
Specifically, we define the 'Brillouin zone filtering' as a mechanism of enhanced TMR in electrode/spacer/electrode tunneling device where spin filtering is provided by (1) existence of so-called 'hot spot' - special part in the in-plane two-dimensional Brillouin zone (2D BZ) where the semiconductor spacer has very high probability of transmission, and (2) absence of states with in-plane momentums $k_{||}$ corresponding to the 'hot spot' in one spin channel of the ferromagnetic electrode and presence of states with corresponding $k_{||}$ in another spin channel (see Fig 1 for illustration).
Since finite area of the 2D BZ is involved, the Brillouin zone spin filtering mechanism potentially leads to exponential increase of TMR with semiconductor thickness, $t$, $TMR \propto exp(2\Delta \gamma t)$, where $\Delta \gamma$ is the difference of the minimal attenuation constants of the spacer in areas of the 2D BZ available for majority and minority electrons of the electrode. This is much stronger dependence compared to linear increase of TMR with semiconductor thickness  originated from   ‘symmetry spin filtering’ where filtering mechanism works only in close vicinity of some high-symmetry points of 2D BZ (e.g. $\Gamma$ point in  conventional Fe/MgO/Fe MTJs   \cite{Butler01,Mathon01}).

\begin{figure}[t]
\includegraphics*[width=8.5cm]{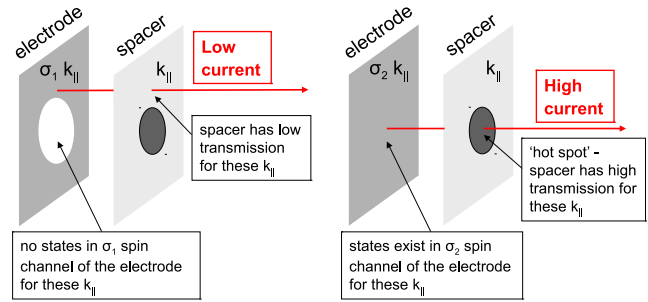}
\caption{(color online). Illustration of the 'Brillouin zone spin filtering' mechanism of enhanced TMR (see text). }
\label{Fig0}
\end{figure}

While the BZ filtering is general   mechanism of enhanced spin dependent transport, in present letter we focus on  novel physics of  realization of the mechanism  in \emph{tunneling} devices.
%based on  low symmetry hcp-Co electrodes and BN spacer.
We show that the BZ filtering conditions can be satisfied in p-doped BN  MTJ with hcp Co electrodes. The many-electron perturbation theory was used to quantify excited electron states and thus find conditions for spin dependent  transport enhanced beyond that of reported for  CoFe/MgO system.
In addition, low crystal symmetry realization of BZ filtering mechanism is of significant potential technological interest in the  context of novel STT-MRAM technology that has a potential to become an 'universal memory' \cite{Akerman05}  combining all the strengths and none of the weaknesses of existing memory types. 

In STT-MRAM devices the perpendicular magnetic anisotropy (PMA) of   electrodes is preferable  option as compared to electrodes with in-plane anisotropy due to faster switching with low current, higher thermal stability and scalability  \cite{Mangin06,Meng06}.
The requirements on PMA electrodes include high thermal stability at reduced dimensions, low switching current   and high TMR all at the same time. Various PMA materials have been studied for STT-MRAM electrodes, including (Co,Fe)/(Pt,Pd) multilayers \cite{Mangin06,Meng06,Mizunama09}, L1$_{0}$-ordered (Co,Fe)Pt  alloys \cite{Kim08,Yoshikawa08,Mizukami11}, rare-earth/transitional-metal (RE/TM) alloys \cite{Hatori07,Nakayama08}, ultra-thin CoFeB \cite{Ikeda10} and CoFeAl \cite{Wang10,Wen11} films, and tetragonal manganese alloys such as Mn$_3$Ga \cite{Kubota12}. Despite of intensive research none of   these   materials   satisfy strict requirements that would allow STT-MRAM to replace conventional memory today. Electrodes with large concentration of heavy Pt or Pd elements (like multilayers Co/Pt or Co/Pd)  have high PMA but also exhibit relatively large Gilbert damping constant  due to strong spin-orbit coupling of Pt and Pd, and,  according to Slonczewski-Berger formula \cite{Slonczewski96, Berger96}, high switching current density.
%Magnetic properties of RE/TM alloys are sensitive to oxidation during fabrication %process, especially for sub-100nm dimensions.
PMA in CoFeB/MgO   \cite{Ikeda10} and in Co$_2$FeAl/MgO MTJs \cite{Wen11}   demonstrated recently with ultra-thin layers of CoFeB and Co$_2$FeAl originates from the electrode/spacer interface, not the volume of the electrodes.
%But switching current increases (due to sharp increase of damping constant) and TMR %degrades as thickness of the CoFeB or Co$_2$FeAl film decreases.
Finally, TMR in Mn$_3$Ga/MgO MTJs was found very small, far below the application range.

In present letter we study BZ filtering mechanism of (exponentially) large TMR in MTJ based on hcp Co(0001) electrodes and  h-BN spacer. h-BN is an ultrahigh chemically stable semiconductor with   band gap of $\sim$6 eV that has   prefect lattice matching  with   hcp Co. Growth of single and multiple layers of h-BN on Co(0001) has been recently  demonstrated \cite{Orofeo11}. To the best of our knowledge only few   studies of Co/BN/Co system exist, mainly with single sheet of BN  spacer \cite{Yazev11,Joshi13,Zhou11}. The FM/BN/Gr/FM(111) MTJ has been   suggested in  \cite{Karpan11}, where  the ferromagnet (FM) like   fcc Ni or Co  do not have PMA . High TMR in this design is achieved by using graphite (Gr) as a spin filter, while overall resistance of MTJ is regulated by thickness of   h-BN. In present work we show that complex BN/Gr structure is not needed and h-BN   itself can act as a spin filter to produce high TMR. Also,    h-BN is more oxidation   and intercalation resistant than graphite.

\begin{figure}[t]
\includegraphics*[width=8.5cm]{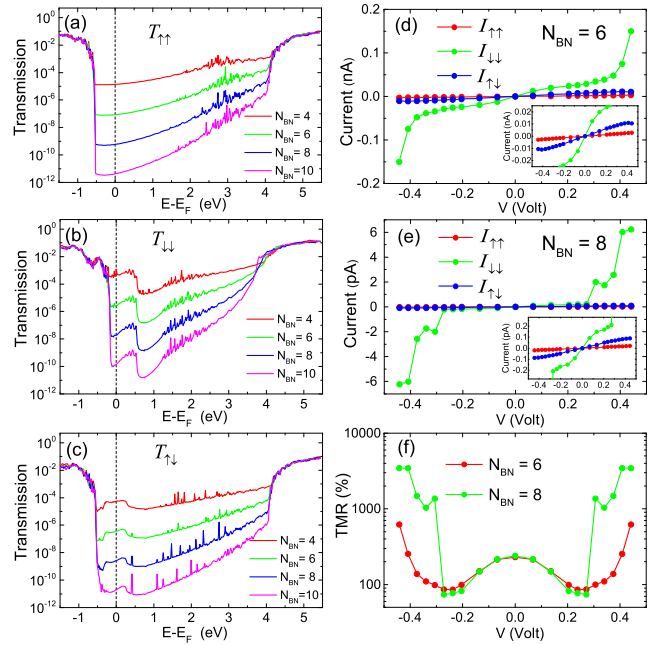}
\caption{(color online). Left panels: transmission functions (a) $T_{\uparrow \uparrow}$, (b) $T_{\downarrow \downarrow}$, and (c) $T_{\uparrow \downarrow}$   calculated at zero  voltage for Co(0001)/h-BN/Co(0001) MTJ with $N_{BN}$ = 4, 6, 8, and 10. Right panels:  currents $I_{\uparrow \uparrow}$, $I_{\downarrow \downarrow}$, and $I_{\uparrow \downarrow}$  for (d) $N_{BN}$ = 6   and (e) $N_{BN}$ = 8, and (f) corresponding TMR     shown as function of bias   voltage. Inserts in (d) and (e) show I-V curves at smaller scale. Note logarithmic scale in   (a), (b), (c)  and (f).}
\label{Fig1}
\end{figure}
We calculated transmission and I-V curves of the Co(0001)/h-BN/Co(0001) MTJ    using nonequilibrium Green's function (NEGF) approach  \cite{Faleev05} developed within the LDA-based LMTO-ASA formalism \cite{Turek97,Schilfgaarde98}. Geometry of the system consist of   $N_{BN}$ h-BN sheets sandwiched between two semi-infinite hcp Co electrodes. The in-plane lattice constant,  2.50 \r{A}, and distances between   Co  layers, 2.03 \r{A}, and  between h-BN layers, 3.33 \r{A}, were set to    experimental values for bulk hcp Co and h-BN. Positions of surface B and N atoms   relative to surface Co layer were calculated by using VASP molecular dynamic program \cite{Kresse96}. In lowest energy configuration  N and B atoms occupy top and hcp sites with distance between Co and BN layers equals to 3.31 \r{A} (see, e.g., \cite{Joshi13} for definition of top, fcc and hcp sites  on the Co(0001) surface). Two configurations (N,B)=(top,hcp) and (N,B)=(top,fcc) have similar energies, in agreement with previous calculations
%for a monolayer of h-BN on hcp Co(0001) surface
\cite{Joshi13,Zhou11} and experimental observations \cite{Orofeo11}.
%Other possible configurations (N,B)=(hcp,top),(hcp,fcc), (fcc,top), and (fcc,hcp) have  slightly higher energy (0.02 to 0.03 eV per unit cell).

Transmission functions $T_{\uparrow \uparrow}$, $T_{\downarrow \downarrow}$, and $T_{\uparrow \downarrow}$ calculated at zero bias voltage for Co(0001)/h-BN/Co(0001) MTJ with $N_{BN}$ = 4, 6, 8, and 10 are shown on left panels of Fig 2. Here  $T_{\uparrow \uparrow}$   and $T_{\downarrow \downarrow}$  are   transmission in  majority and minority channels calculated for parallel magnetization configuration of electrodes and $T_{\uparrow \downarrow}$   is  transmission   calculated for antiparallel   configuration. The I-V curves of $I_{\uparrow \uparrow}$, $I_{\downarrow \downarrow}$, and $I_{\uparrow \downarrow}$ currents per unit cell area are shown on Fig 2(d) and Fig 2(e) for $N_{BN}$ = 6   and   8. Corresponding   $TMR = (I_{\uparrow \uparrow} + I_{\downarrow \downarrow} - 2I_{\uparrow \downarrow})/2I_{\uparrow \downarrow}$ is shown on Fig 2(f) as function of bias  voltage, $V$. As seen on Fig 2(d),(e) the $I_{\downarrow \downarrow}$  at small voltages  is about 7 times larger compared to $I_{\uparrow \downarrow}$ (and even more so for $I_{\uparrow \uparrow}$)  resulting in $\sim 250 \%$ TMR at $V< 0.1$ V [see Fig 2(f)]. This is mostly due to the fact that density of states (DOS) of minority electrons of hcp Co at Fermi energy, $E_F$,   is significantly larger than  DOS of majority electrons [see Fig 3(a) and Fig 3(b)].
%for DOS  of majority and minority electrons of hcp Co at $E_F$ averaged over momentum along $c$-axis, $k_z$, as function of the in-plane momentum $k_{||}$ in 2D Brillouin zone (BZ)].

\begin{figure}[h]
\includegraphics*[height=3.5cm]{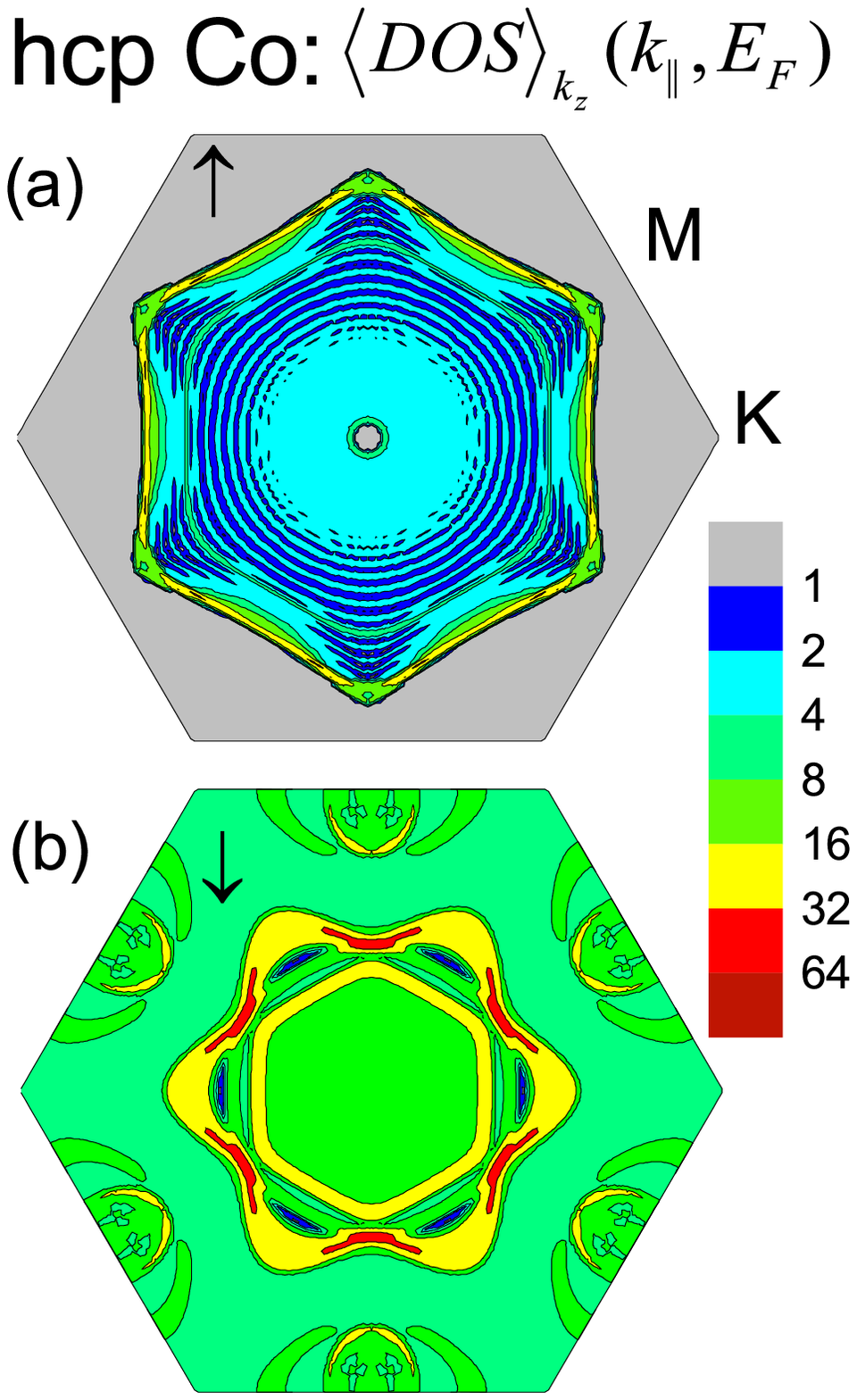}\
\includegraphics*[height=3.5cm]{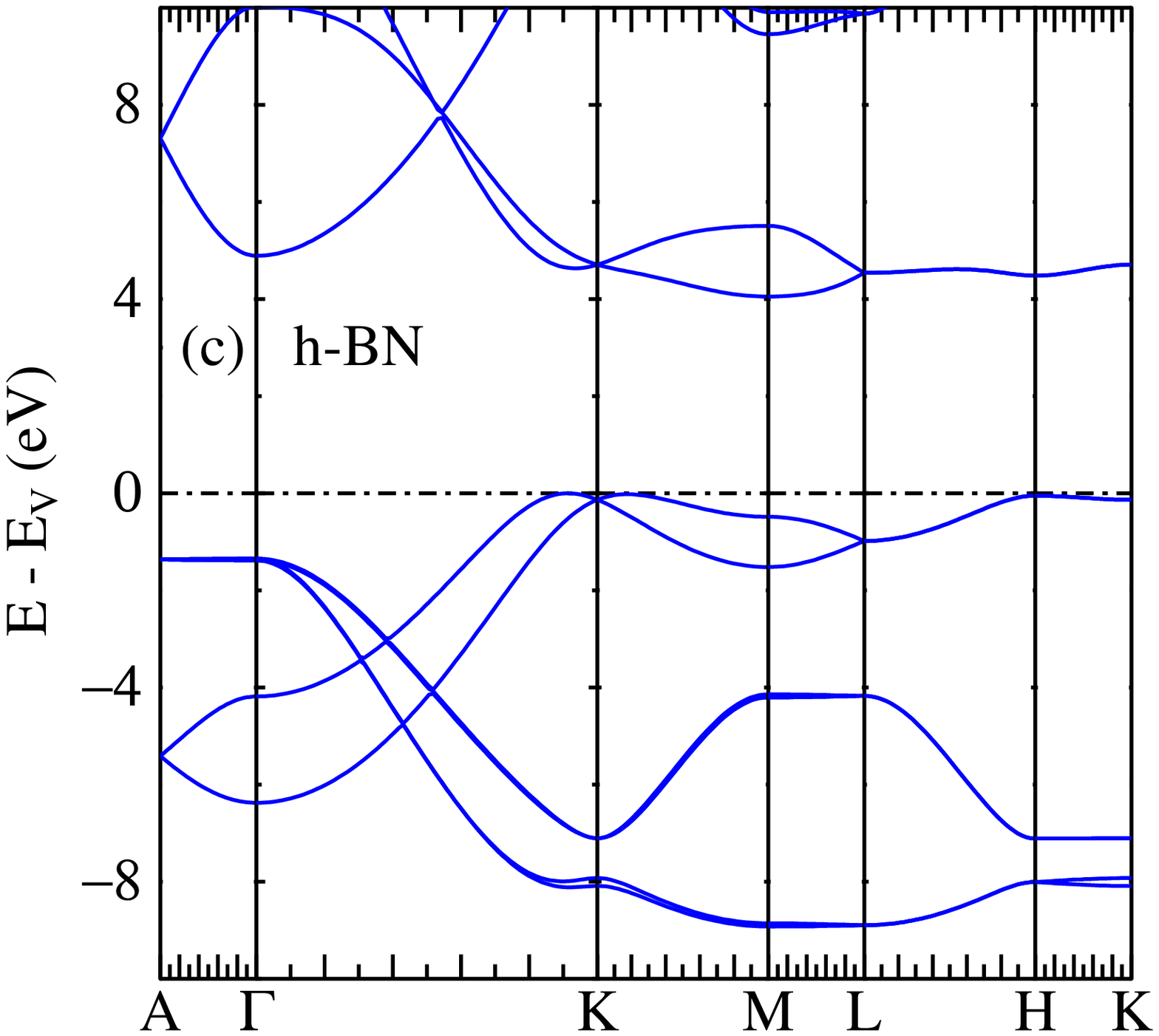}\
\includegraphics*[height=3.5cm]{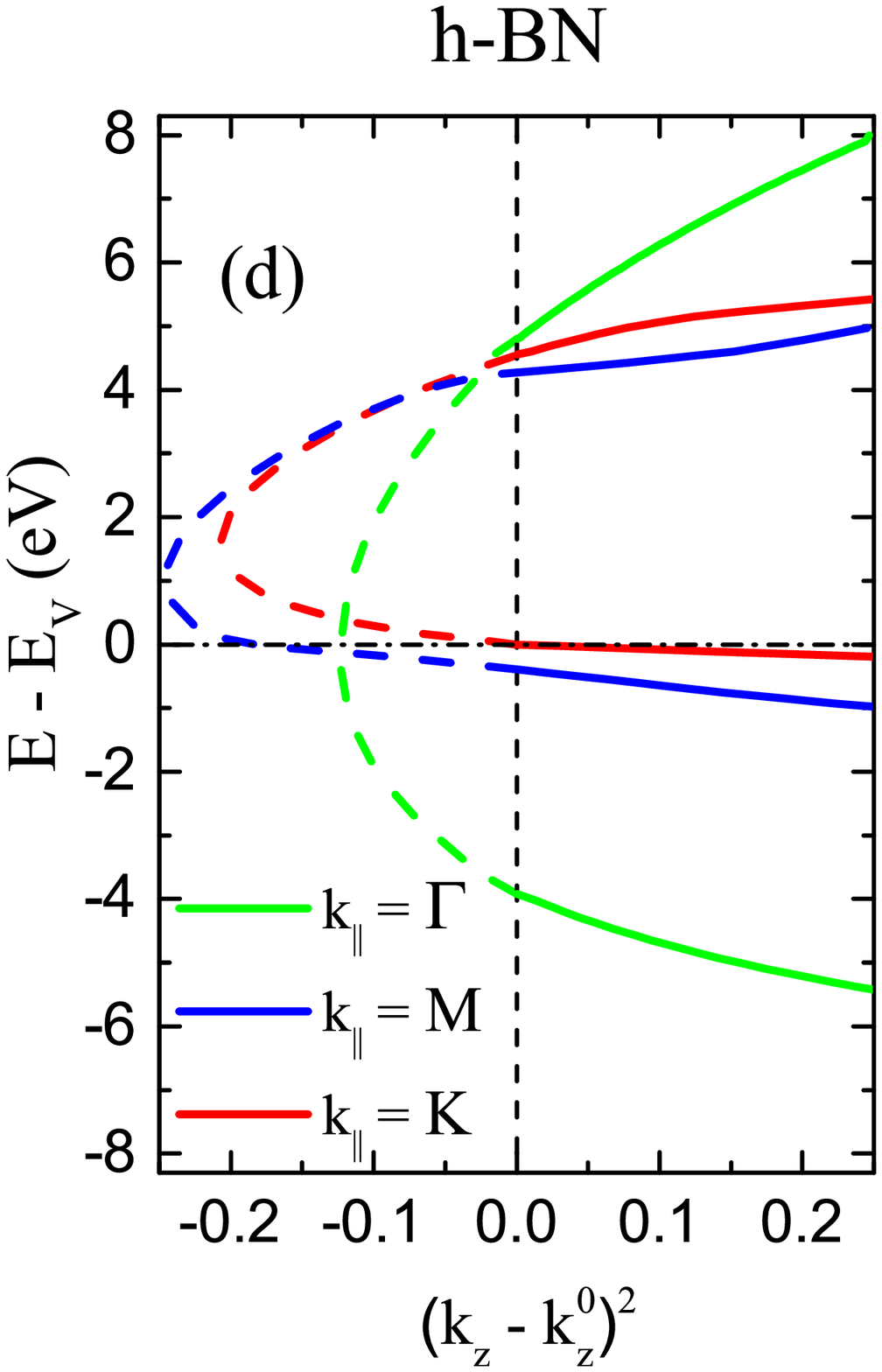}
\caption{(color online).   (a) majority and (b) minority electrons DOS of bulk hcp Co at $E_F$
averaged over the wave number along $c$-axis, $k_z$, shown as function of  in-plane $k_{||}$ in 2D Brillouin zone (BZ). Note absence of majority electron states near   $K$  and $M$ points.
%averaged over  $k_z$ shown as function of  $k_{||}$ in 2D BZ (note logarithmic color scale, arbitrary units).
(c) bands of bulk h-BN calculated by   full potential (FP) LMTO method. (d) selected complex bands of h-BN calculated by Green's function approach within LMTO-ASA   \cite{Turek97,Schilfgaarde98} shown  at three $k_{||}=\Gamma, M$ and $K$ as function of $(k_z - k_z^0)^2$, where $k_z^0$ is the wave number corresponding to the lowest energy of conduction band at given $k_{||}$ [$k_z^0=0$ for  $k_{||}=\Gamma$ and $M$, and $k_z^0=0.5$ for  $k_{||}=K$]. Negative $(k_z - k_z^0)^2$
%shown by dashed lines
represent evanescent modes
with $k_z(E)=k_z^0 + i\gamma(E)$.
%[Due to ASA approximation bands somewhat more dispersive than FP-LMTO bands.]
}
\label{Fig2}
\end{figure}
More interesting, however, is the sharp increase of $I_{\downarrow \downarrow}$ and TMR that occurs at   $V\approx 0.4$ V for $N_{BN}$ = 6 and at $V\approx 0.3$ V for $N_{BN}$ = 8 (TMR is as high as 4000$\%$ at $0.4$ V  for $N_{BN}$ = 8).  Three features of transmission functions $T_{\uparrow \uparrow}$, $T_{\downarrow \downarrow}$, and $T_{\uparrow \downarrow}$  are essential for  understanding high TMR of the system predicted by LDA   at $V\ge 0.3 $ V. These   features are (1)   sharp drop of transmission functions at energy near the valence band maximum ($E_V$) of the h-BN slab, (2) alignment of the Fermi energy,  $E_F$,   of   Co  inside the h-BN band gap very close to $E_V$ (LDA predicts $\Delta E_{FV} \equiv E_F - E_V \approx 0.2$ eV), and (3) sharp drop of $T_{\downarrow \downarrow}$  occurs at   higher energy compared to energy where the sharp drop of $T_{\uparrow \uparrow}$ and $T_{\uparrow \downarrow}$   occurs [as  seen on Fig 2     $T_{\downarrow \downarrow}$ is several orders of magnitude (!) larger than $T_{\uparrow \downarrow}$ or $T_{\uparrow \uparrow}$ in   energy window $E_F-0.6$ eV $< E < E_F-0.2$ eV]. The last feature is a consequence of the BZ filtering as will be described below in discussion of Fig 4.
%Assuming that above features of equilibrium transmissions do not change much when finite bias voltage is applied and that energy integration window for calculation of currents can be approximated by ($E_F$ – eV/2:$E_F$ + eV/2), the sharp rize in $I_{\downarrow \downarrow}$ and TMR is expected to happen at V $\sim 2\Delta_{FV}/e \sim 0.4 V$, in agreement with results of NEGF calculations.

Let us first explain the sharp drop of transmission functions shown on Fig 2(a)-(c) at energy near $E_V$.
% (note that transmission functions of conventional Fe/MgO/Fe MTJs    do not have sharp features for energies near the MgO band gap edges).
On Fig 3(c) and Fig 3(d) we show real bands and  selected complex bands of h-BN. In bulk h-BN the
mode with smallest attenuation constant $\gamma(E) = Im{(k_z)}$ [shown by dashed green line in Fig 3(d)] is the mode with $k_{||}=0$ for all energies, $E$, inside the band gap except energies in close vicinity of valence band maximum, $E_V$, or conduction band minimum, $E_C$.
%smallest attenuation constant $\gamma(E) = Im{(k_z)}$ of evanescent modes propagating in $z$ direction with   energy $E$ inside the band gap corresponds to the mode with $k_{||}=0$,  with exception of $E$ in close vicinity of $E_V$ and conduction band minimum, $E_C$.
This evanescent mode
%[shown by dashed green line in Fig 3(d)] has $\Delta_1$ symmetry and
continuously connects   two real h-BN bands
%with $\Delta_1$ symmetry
shown along $A-\Gamma$ line on Fig 3(c):   highly dispersive conduction band with energy $E_V + 4.8$ eV at $\Gamma$ point and   valence band with energy $E_V - 4.1$ eV at $\Gamma$ point. As seen on Fig 3(d) maximum of this $\gamma(E)$ occurs at $E\sim E_V$.
%in the middle of the energy range ($E_V - 4.1$ eV: $E_V + 4.8$ eV) closer to its lower edge.
This explains why transmission functions shown in Fig 2 have a tendency to smoothly decrease (on logarithmic scale) when energy decreases from $E_C$ to $E_V$ with a minimum near $E_V$.
%(we ignore here other factors that contribute to $E$-dependence of transmission, such as DOS of electrodes, etc).
On the other hand, since the valence band maximum (VBM) of h-BN corresponds to a state with momentum near the $K$ point, for $E$ slightly above $E_V$ the  evanescent mode with smallest attenuation constant [shown by dashed red line in Fig 3(d)] is the mode with $k_{||}$ close to the $K$ point. Since the valence band along the $H-K$ line with energy near $E_V$ is almost flat [see Fig 3(c),(d)], corresponding $\gamma(E)$ increases very fast when  $E$ increases above $E_V$. Thus the behavior of the smallest attenuation constant of BN  inside the band gap
%combination of the small mass (in $z$-direction) of the lowest-energy conduction   band of bulk h-BN with $k_{||}$ at $\Gamma$ point and very large  mass (in $z$-direction)  of the highest-energy valence band with $k_{||}$ at $K$ point
explains general shape  of   transmission functions of Co/h-BN/Co MTJ for $E$ inside the h-BN band gap and  their sharp drop near $E_V$. In other words, specific features of the complex band structure of BN described above are responsible for ultra-sensitivity of spin-dependent conductance and TMR of this MTJ to position of the Fermi energy inside the BN band gap.
(Note that transmission functions of conventional MgO-based MTJs do not exhibit sharp features as function of energy since MgO has much simpler complex band structure as compared to that of h-BN.)

%In MgO the evanescent mode with smallest attenuation constant is the mode with $k_{||}=0$ for all energies inside the MgO band gap. This mode continuously connects two highly dispersive bands that have $E=E_C$ and $E=E_V$ at $k_{||}=0$, therefore smallest $\gamma(E)$ smoothly vanishes near the band gap edges and transmission functions of MgO-based MTJs do not have sharp features near $E_C$ or $E_V$.)
%%[As opposed to h-BN the smallest $\gamma(E)$ of conventional MgO spacer and

%For comparison, smallest $\gamma(E)$ of conventional MgO spacer   vanishes at band gap edges and corresponding transmission functions do not have sharp features near $E_C$ or $E_V$.

%Let us explain why sharp drop of $T_{\uparrow \uparrow}$ and $T_{\downarrow \downarrow}$  in Fig 1(a) and Fig 1(b) occur at different energies.
Left panels of Fig 4    show   bands of repeated slabs of 9 hcp Co and 5 h-BN layers calculated along symmetry lines of 2D BZ  by LDA-based FP-LMTO method.
%\cite{Kotani02,Kotani07}.
Green and blue colors of bands on Fig 4 are mixed according to the values of projections    of the wave functions to Co and BN atomic orbitals, correspondingly.
Fig 4(a),(d) show  that  highest energy at which majority Co states and valence h-BN states co-exist at the same $k_{||}$ is $E'_V= E_F-0.6$ eV, while highest energy at which minority Co  states and valence h-BN states co-exist is the VBM of h-BN $E_V=E_F - 0.2$ eV. This explains why the sharp drop of $T_{\downarrow \downarrow}$ and $T_{\uparrow \uparrow}$ (together with $T_{\uparrow \downarrow}$) shown on Fig 2 occur at two different energies: $E_F - 0.2$ eV and $E_F - 0.6$ eV, correspondingly.

Let us stress again that it is the BZ filtering mechanism which is responsible for several orders of magnitude difference of $T_{\downarrow \downarrow}$ transmission compared to $T_{\uparrow \uparrow}$ (and $T_{\uparrow \downarrow}$) at energies in the range ($E_F - 0.2$ eV, $E_F - 0.6$ eV). The real states of BN (or 'hot spot') near the $K$ point at these energies provide metallic transmission for minority Co electrons, $T_{\downarrow \downarrow}$, while absence of majority Co states in vicinity of $K$ point at these energies makes $T_{\uparrow \uparrow}$ and $T_{\uparrow \downarrow}$ exponentially suppressed.

%Fig 3(a) shows that  largest energy, $E'_V$, at which majority Co states and valence h-BN states co-exist at the same $k_{||}$ is $E'_V= E_F-0.6$ eV.
%, that is less than VBM of h-BN ($E_V=E_F - 0.2$ eV).

%Fig 3(d) shows that largest energy at which minority Co  states and valence h-BN states co-exist is the VBM of h-BN ($E_V=E_F - 0.2$ eV). Thus, sharp drop of $T_{\downarrow \downarrow}$ seen on Fig 1 occurs at $E_V$, while sharp drop of
%$T_{\uparrow \uparrow}$ and $T_{\uparrow \downarrow}$ occurs at smaller energy, $E'_V$, as a consequence of the fact that majority electrons of  hcp Co do not have states at $k_{||}$ near $K$ point of 2D BZ at $E$ near $E_F$ [see Fig 2(a) and Fig 3(a)], while VBM of h-BN occurs near the $K$ point [see Fig 2(c) and Fig 3(a),(d)].

\begin{figure}[t]
\includegraphics*[width=2.7cm]{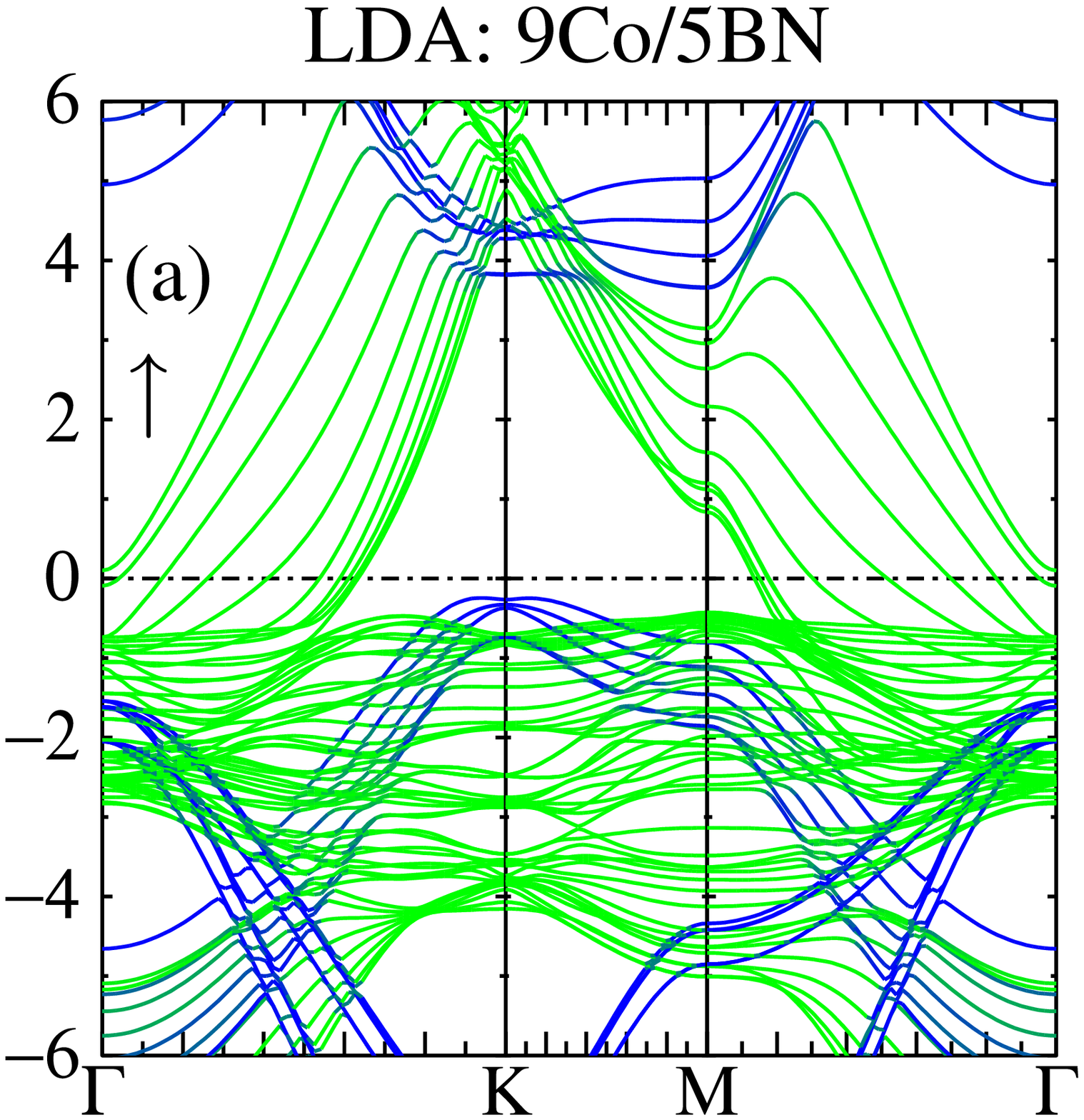}\
\includegraphics*[width=2.7cm]{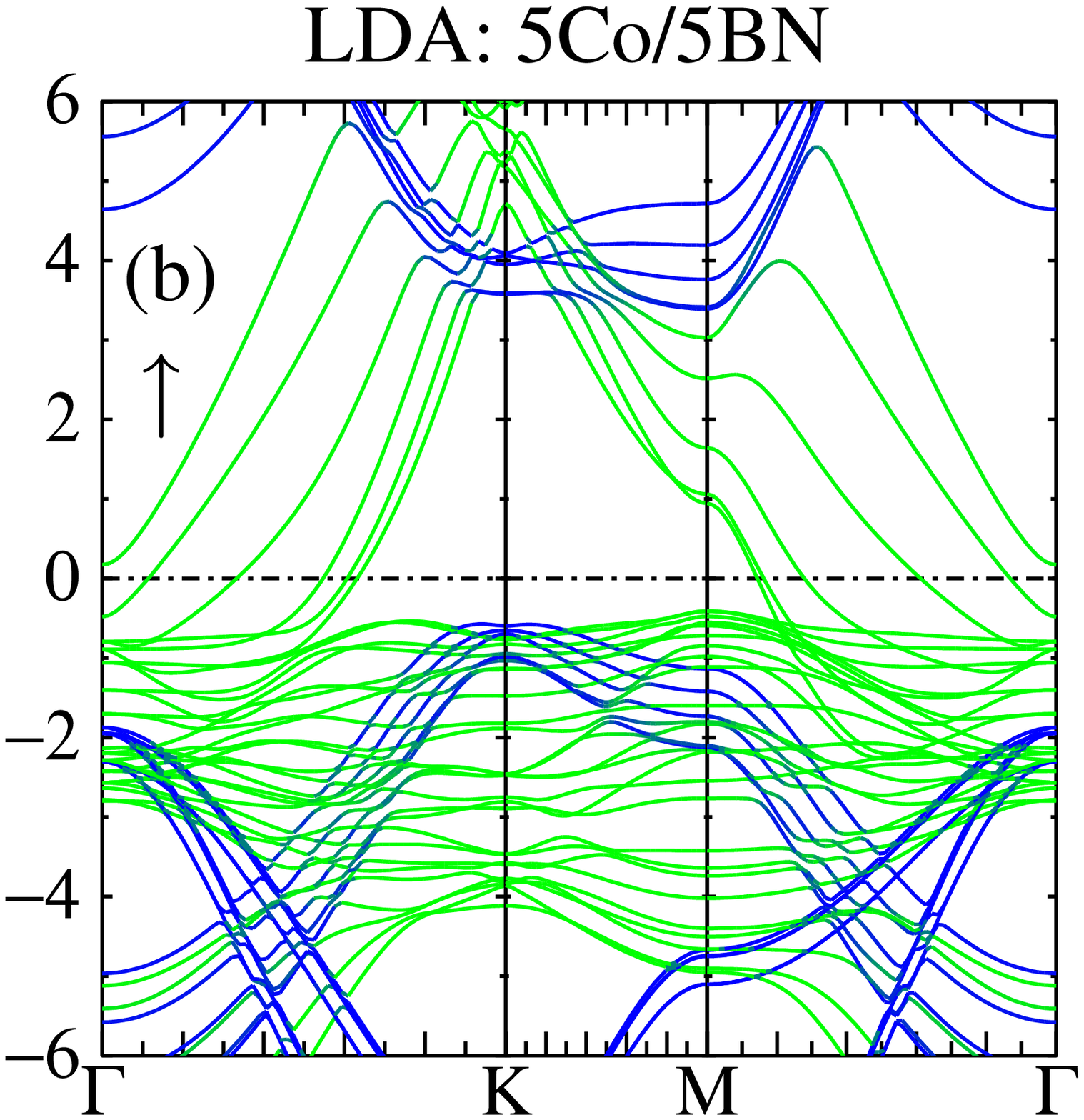}\
\includegraphics*[width=2.7cm]{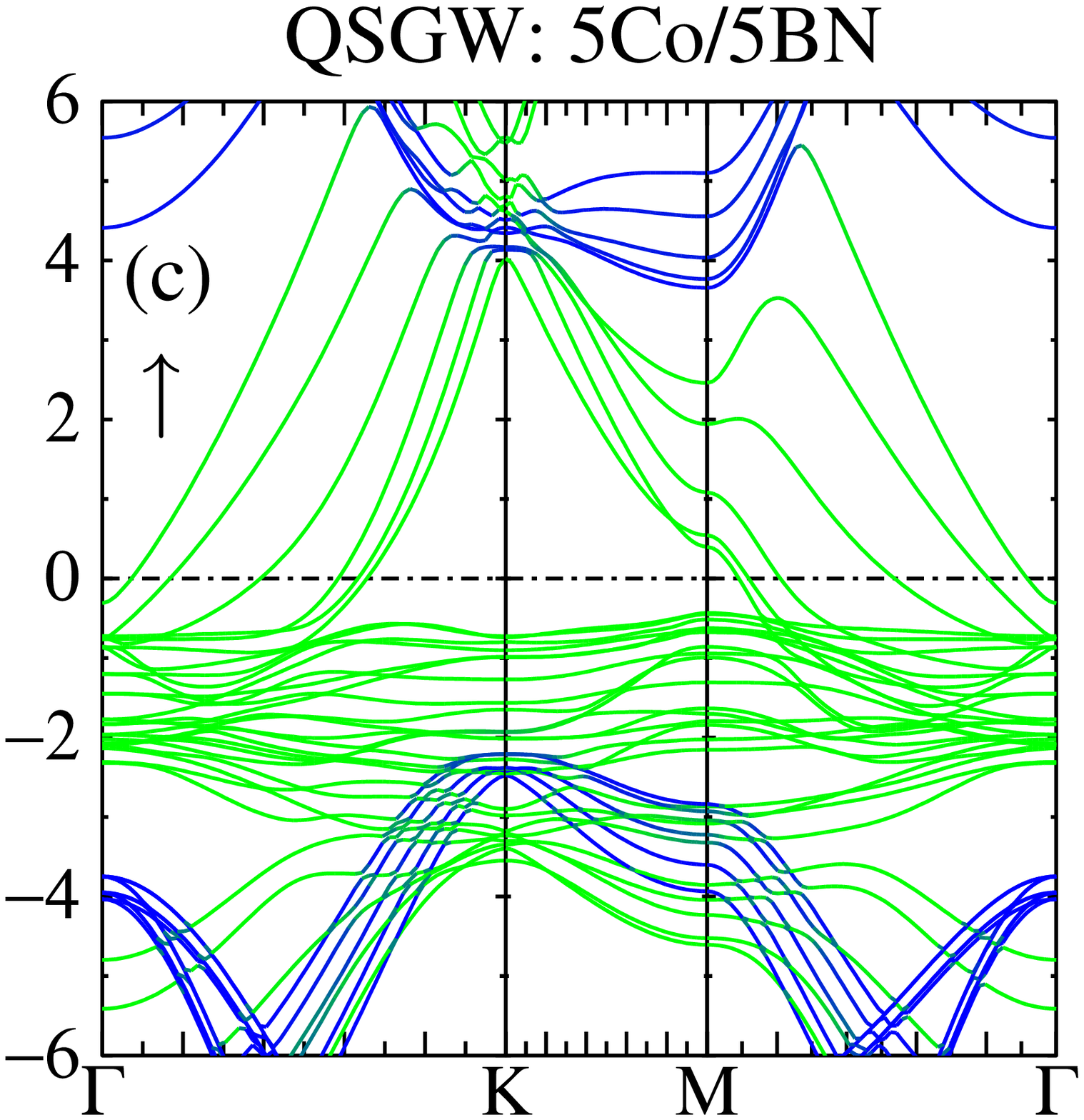}
\includegraphics*[width=2.7cm]{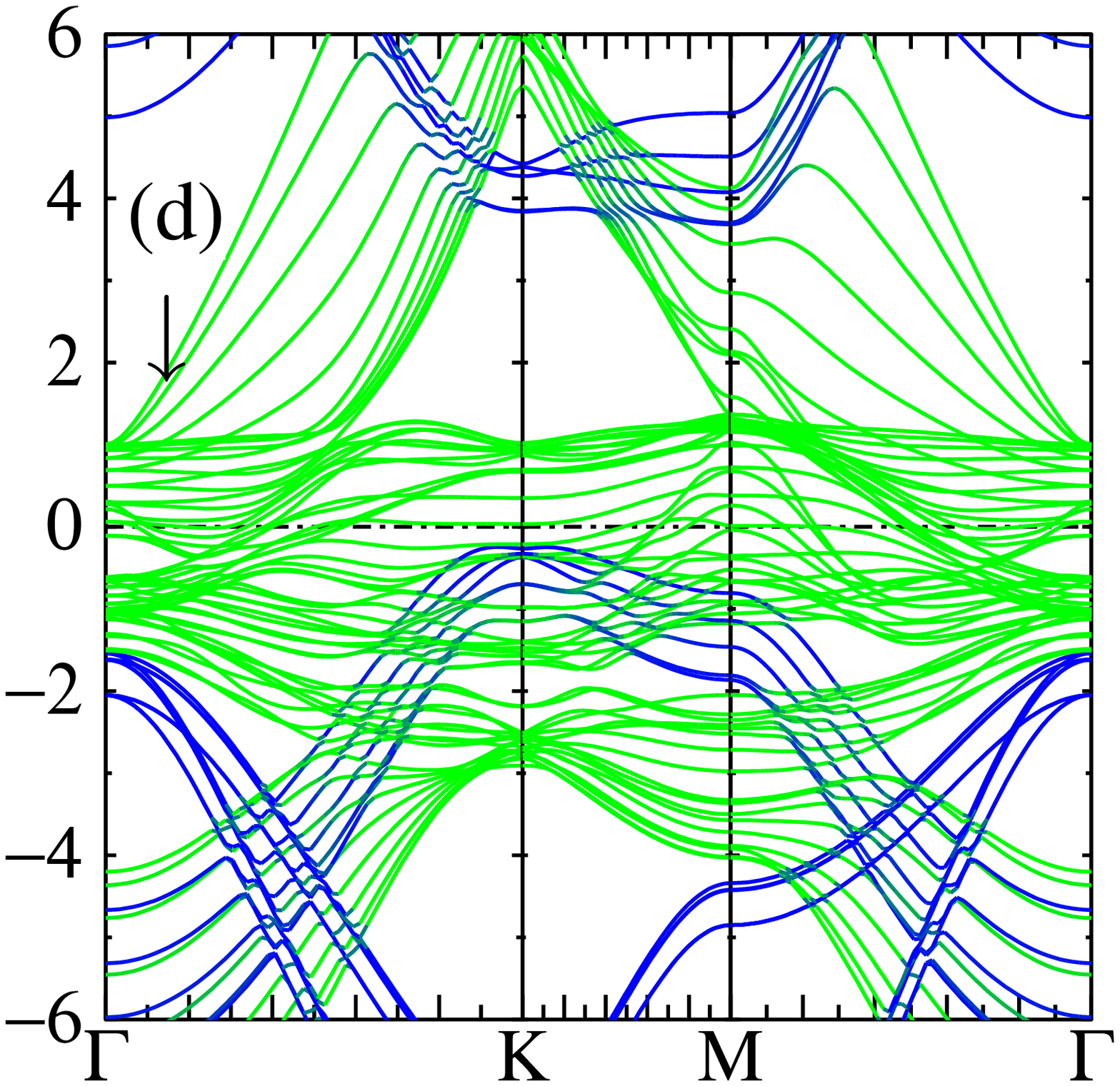}\
\includegraphics*[width=2.7cm]{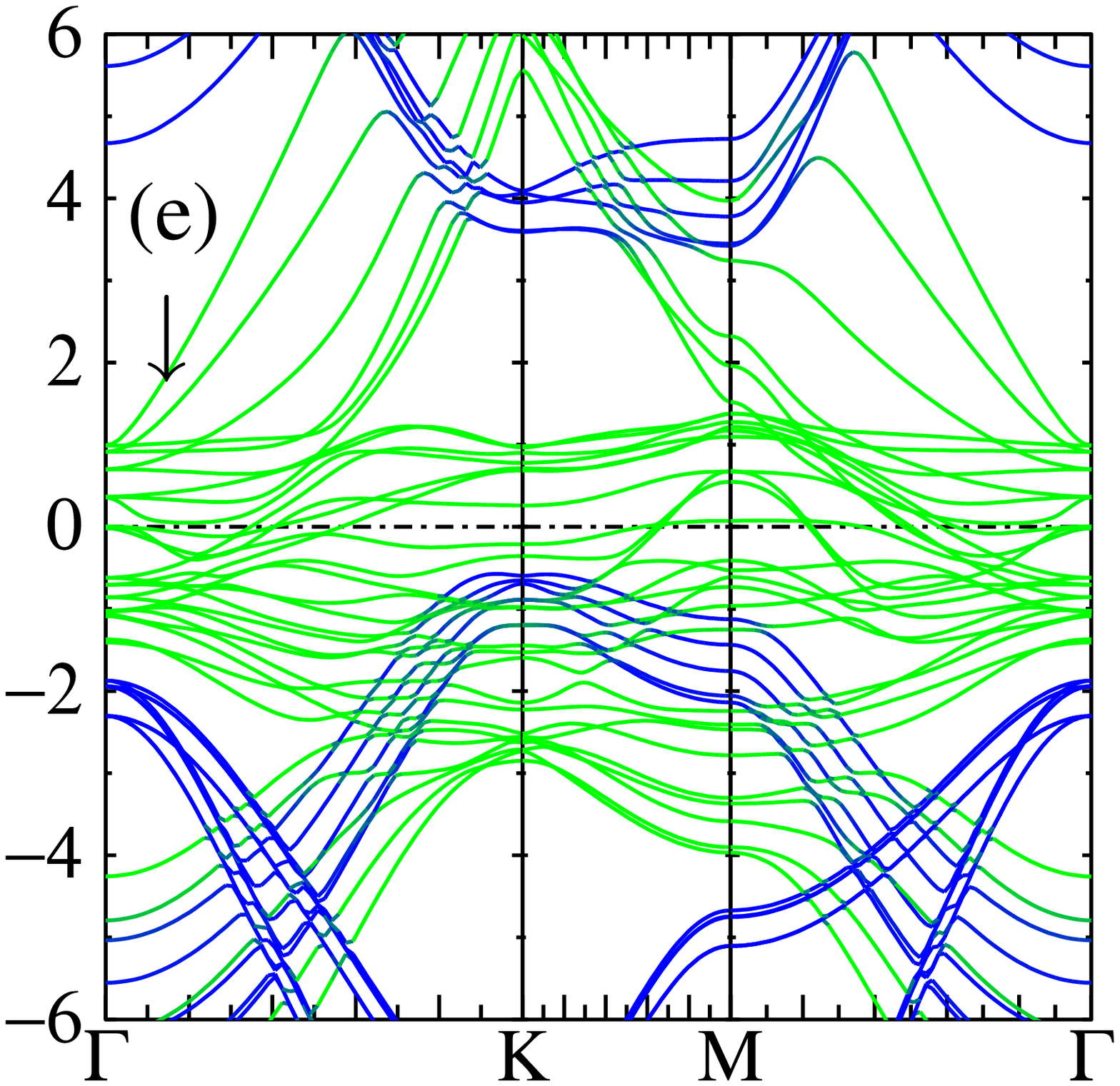}\
\includegraphics*[width=2.7cm]{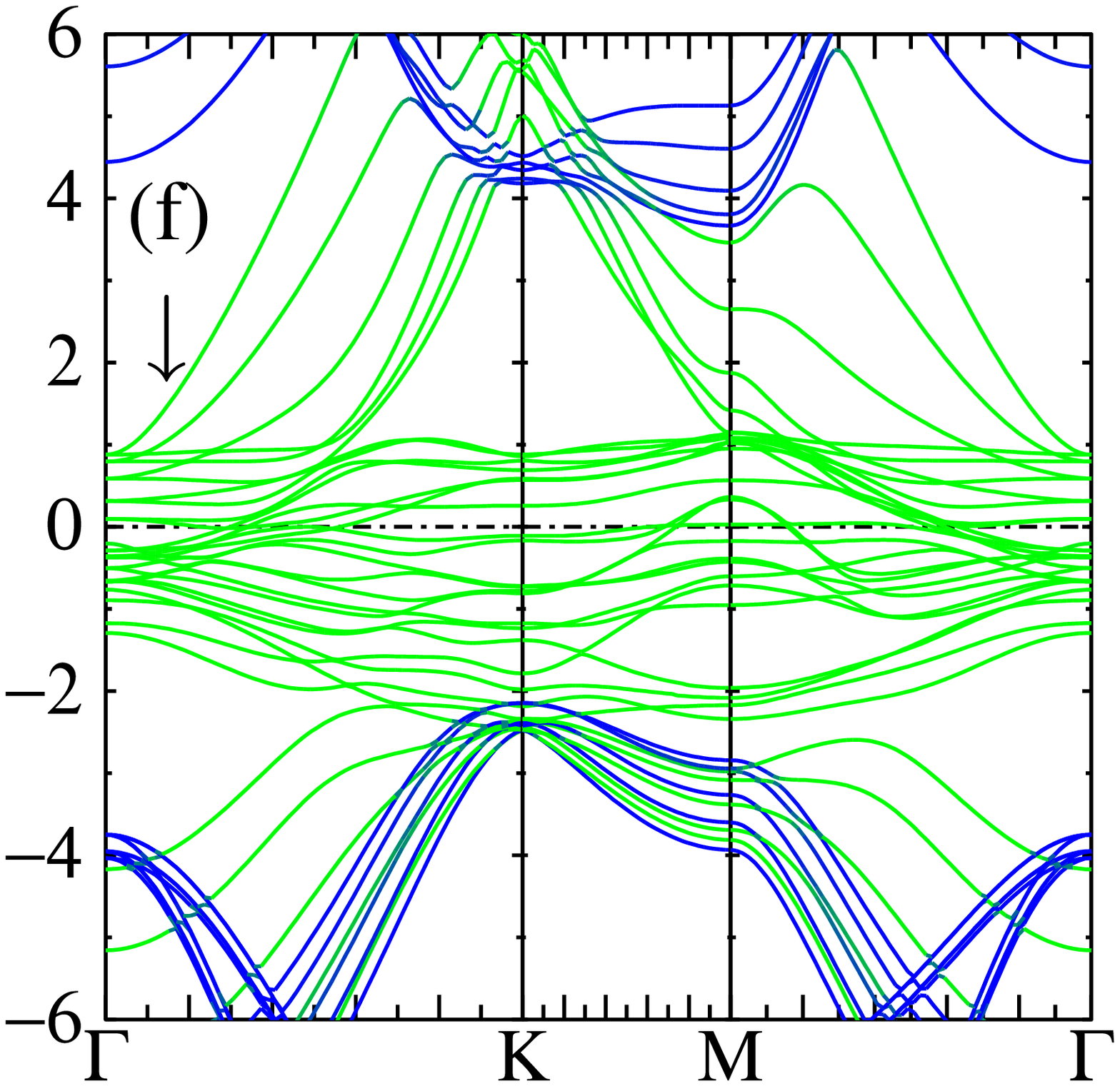}
\caption{(color online).   Top panels: majority    bands of repeated slabs of (a) 9Co/5BN and (b) 5Co/5BN  calculated using the LDA theory, and (c) 5Co/5BN   calculated using the QSGW theory (y-axis on all graphs is $E-E_F$ in eV). Bottom panels: corresponding minority bands.
%See text for color description.
%Green and blue colors of bands are mixed in accordance to the projection weights of the wave function to Co and BN atomic orbitals, correspondingly.
}
\label{Fig3}
\end{figure}
Our NEGF calculations show that for small   bias voltage, $V$,  applied to Co/BN/Co system the equilibrium transmissions functions shown on Fig 2 do not change much if $E_F$ is replaced by average $\langle E_F \rangle = (E_F^L + E_F^R)/2$, where $E_F^L$ and $E_F^R$ are Fermi energies of the left and right electrodes ($E_F^R - E_F^L = eV$). Thus, the onset of sharp rise of $I_{\downarrow \downarrow}$ could be estimated at   $V\approx 2\Delta E_{FV}/e \approx 0.4$ V [that  agrees with actual NEGF onset at $0.3-0.4$ V shown on Fig 2(d),(e)] and onset of sharp rise of $I_{\downarrow \uparrow}$ and $I_{\uparrow \uparrow}$ could be estimated at   $V\approx 2(E_F - E'_V)/e \approx 1.2$ V, which is beyond the scale of voltages shown on Fig 2.
%Therefore the voltage window in which LDA predicts large TMR could be estimated as ($0.3 - 1.2$) V.

%Let us assume that for small   bias voltage, $V$,  applied to Co/BN/Co system the equilibrium transmissions functions shown on Fig 1 do not change much, with $E_F$ being replaced by average $\langle E_F \rangle = (E_F^L + E_F^R)/2$, where $E_F^L$ and $E_F^L$ are Fermi energies of the left and right electrodes, so $E_F^R - E_F^L = eV$. Then  the sharp rise of $I_{\downarrow \downarrow}$ would occur at $V\approx 2(E_F - E_V)/e \approx 0.4$ V, in agreement with results of NEGF calculations shown on Fig 1(d) and Fig 1(e). Analogously, the sharp rise in $I_{\uparrow \uparrow}$ and $I_{\uparrow \downarrow}$ would occur at $V\approx 2(E_F - E'_V)/e \approx 1.2$ V resulting in large TMR in the voltage window (0.4 $:$ 1.2) V.

%Since spin filtering due to high transmission through real h-BN states can only occur at energies near $E_V$
%The high TMR shown on Fig 1(f) at $V> 0.3$ V is very sensitive to the Co Fermi energy being aligned close to the VBM of h-BN.
Our results are very sensitive to the position of the Fermi energy relative to the bottom edge of the BN band gap, $\Delta E_{FV}$, since   onset of high TMR occurs at voltage $V\simeq 2 \Delta E_{FV}/e$, yet $V$ cannot exceed the electrical breakdown limit.
%The larger is the difference $\Delta E_{FV}$ the larger voltage should be applied to obtain high TMR. On the other hand, applying voltage larger than 0.4 V is unpractical.
We studied dependence of   $\Delta E_{FV}$ on variations in geometry of the systems by  performing FP-LMTO calculations within LDA   for repeated slabs of $N_{BN}$ h-BN   and $N_{Co}$ hcp Co layers (varying $N_{BN}$ and $N_{Co}$ from 4 to 12) for all  configurations of B and N atoms (B,N)=(top,hcp), (top,fcc), (hcp,top), (hcp,fcc), (fcc,top), and (fcc,hcp) and for reasonable variations of positions of surface B, N and Co atoms. The value of $\Delta E_{FV}$  was found to be  stable within LDA/DFT, $\Delta E_{FV}\approx 0.2$ eV, in all above cases for sufficiently thick Co slab with $N_{Co}\ge 7$.
%(for 5 Co layers the difference slightly increases to $\Delta E_{FV} \approx 0.5$).

On the other hand, it is well known that LDA underestimates   band gaps of semiconductors  and poorly describes the Schottky barrier heights \cite{Das89}. To evaluate $\Delta E_{FV}$ on the level beyond LDA  we performed calculations for periodic slab of 5 hcp Co  and 5 h-BN layers by using the QSGW theory that is known to describes band gaps and other properties of materials with moderate e-e correlations significantly better than LDA \cite{Faleev04,Schilfgaarde06,Kotani07}. (The restriction to 5Co/5BN slab size was due to heavy computational costs of QSGW.)
%The QSGW theory has been tested for a wide variety of bulk material systems and has been shown to be a good predictor of materials properties for many classes of compounds composed of elements throughout the periodic table .
Fig 4 shows the band structure of majority and minority electrons calculated for periodic 9Co/5BN  and 5Co/5BN  slabs within the LDA theory, and for 5Co/5BN  slabs within the QSGW theory. It is seen that reducing the size of Co slab from 9 to 5 layers leads to   moderate increase of   LDA $\Delta E_{FV}$ from 0.2 eV to 0.5 eV (as mentioned above $\Delta E_{FV}$ converges to $\approx 0.2$ eV starting with $N_{Co}\ge 7$), whereas more accurate inclusion of the e-e correlations within the QSGW theory results in dramatic increase of $\Delta E_{FV}$ to as large as 2.2 eV. Thus, taking into account correlations on beyond LDA level is very important for accurate prediction of the Fermi energy alignment $(\Delta E_{FV})$ at Co/h-BN interface and, consequently, for designing the Co/h-BN/Co MTJ with high  TMR.

In order to decrease  $\Delta E_{FV}$ from 2.2 eV predicted by QSGW  to practical levels one can p-dope the h-BN (e.g. by Mg).  It was recently shown that Mg acceptor level in Mg-doped h-BN (h-BN:Mg) is as low as 0.031 eV \cite{Dehal11}, so varying the Mg concentration could bring $\Delta E_{FV}$ to desired range of  $\Delta E_{FV}\le 0.1$ eV. Note that for ideal alignment of  $\Delta E_{FV}=0$ the  TMR could be as high as several orders of magnitude at  very low voltages [see Fig 2].

%By analogy with the term 'symmetry filtering'   used to describe the mechanism of enhanced TMR in Fe/MgO/Fe MTJs \cite{Butler08} we can introduce the term 'Brillouin zone filtering' to describe the mechanism of enhanced TMR in Co/h-BN/Co system studied in this work. Specifically, 'Brillouin zone filtering' could be defined as a mechanism of enhanced TMR in FM/spacer/FM devices where spin filtering is provided by absence of states in one spin channel and presence of states in another spin channel of ferromagnetic (FM) electrode in specific part of 2D BZ (common for FM and spacer) where spacer has very high probability of transmission. Note that high  TMR of FM/BN/Gr/FM and FM/Gr/FM MTJs suggested in \cite{Karpan11} also originates from Brillouin zone filtering due to Fermi surface  of graphite   being  single $K$ point in 2D BZ [common for Gr and FM(111)].

In conclusion, general 'Brillouin zone spin filtering'  mechanism of enhanced TMR is described for  magnetic tunnel junctions  and studied on an example of MTJ with hcp Co electrodes and h-BN spacer.
We suggest   new  MTJ for STT-MRAM and STO  applications based on   hcp Co-Pt or Co-Pd disordered alloy electrodes and p-doped h-BN spacer. %One advantage of this choice of materials for MTJ is that
PMA of  such low symmetry hcp electrodes  originates from the whole volume rather than interfaces and does not require chemical ordering at atomic scale promoting small device-to-device variations, a property important for high yield and low cost. Concentration of heavy Pt or Pd atoms could be balanced between requirements of strong PMA   and small damping constant of the alloy (damping constant of pure hcp Co is small, $\alpha_{hcp Co} = 0.004$ \cite{Bhagat74} at 300K).

Owning to the specific complex band structure of the h-BN the spin-dependent tunneling conductance of the system is ultra-sensitive to small variations of the Fermi energy position inside the BN band gap. Our LDA-based NEGF calculations for  Co(0001)/h-BN/Co(0001) MTJ shows high  TMR  at $V \ge 0.3$ V due to Brillouin zone filtering mechanism. Critical property needed for high TMR in this MTJ is an alignment of Co $E_F$ very close to VBM of h-BN. LDA       predicts $\Delta E_{FV}\approx 0.2$ eV, while  $\Delta E_{FV}$ calculated by   more accurate  QSGW theory increases to 2.2 eV. Thus   p-doping (e.g. by Mg) of the  h-BN is needed to reduce $\Delta E_{FV}$ to practical range of below 0.1 eV to ensure high TMR (that can be as high as several orders of magnitude) at low voltages. By varying the Mg concentration (and thus $\Delta E_{FV}$) and h-BN  slab thickness one can independently regulate the TMR and overall resistance of the MTJ.

%Conclusion: advantages of BN spacer: high TMR and high RA regulated by Mg impurity concentration. Compared to graphite space - less complex, impurity and intercall resistant.. advantages of hcp Co - bulk PMA due to hcp, regulated PMA vs damping by Pt/Pd concentration..
% regulate RA and TMR

%\section{Acknowledgement}

S.F. and O.N.M acknowledge the CNMS User support by Oak Ridge National Laboratory
Division of Scientific User facilities.
O.N.M acknowledge partial support  by C-SPIN, one of the six centers of STARnet,
a Semiconductor Research Corporation program, sponsored by MARCO and DARPA.
S.F. would like to thank Ivan Knez and   Barbara Jones for useful discussions, O.N.M. would like to thank Prof. J.P.Wang and D.Mazumdar
for stimulating discussions.
%, Office of Basic Energy Sciences, U.S. Department of Energy.

%\section{References}

\end{document}